\documentclass[conference]{IEEEtran}

\usepackage[cmex10]{amsmath}
\usepackage{tikz}
\usetikzlibrary{arrows,shapes,snakes,positioning,fit,backgrounds,mindmap}
\usepackage[customcolors]{hf-tikz}
\usepackage{colortbl}
\usepackage{hyphenat}
\usepackage{subfigure}
\usepackage[utf8]{inputenc}

\tikzset{
    >=stealth',
    node/.style={
           rectangle,
           rounded corners,
           draw=black!80, very thick,
           text centered
    },
    env/.style={
           fill=blue!20,
    },
    tf/.style={
           fill=orange!60, 
    },
    wf/.style={
           fill=blue!40,
    },
    pil/.style={
           ->,
           thick,
           shorten <=2pt,
           shorten >=2pt},
    inactive/.style={
          opacity=.3,
          dotted
    },
    inactive-pipeline/.style={
          opacity=.2
    }
}

\long\def\/*#1*/{} 

\begin{document}
%
\title{WaveFlow -- Towards Integration of Ultrasound Processing
with Deep Learning}

\author{
\IEEEauthorblockN{
Piotr Jarosik\IEEEauthorrefmark{1}, 
Michal Byra\IEEEauthorrefmark{2},
Marcin Lewandowski\IEEEauthorrefmark{3}\IEEEauthorrefmark{4}}
\IEEEauthorblockA{
\IEEEauthorrefmark{1}Department of Information and Computational Science,\\
\IEEEauthorrefmark{2}Department of Ultrasound,\\
\IEEEauthorrefmark{3}Laboratory of Professional Electronics,\\
Institute of Fundamental Technological Research PAS,\\ 
Warsaw, Poland;\\
\IEEEauthorrefmark{4}us4us Ltd., Warsaw, Poland;\\
email: pjarosik@ippt.pan.pl, mbyra@ippt.pan.pl, mlew@ippt.pan.pl}}

\maketitle

\begin{abstract}
The ultimate goal of this work is a real-time processing framework for ultrasound image reconstruction augmented with machine learning. To attain this, we have implemented WaveFlow -- a set of ultrasound data acquisition and processing tools for TensorFlow. WaveFlow includes: ultrasound Environments (connection points between the input raw ultrasound data source and TensorFlow) and signal processing Operators (ops) library. Raw data can be processed in real-time using algorithms available both in TensorFlow and WaveFlow. Currently, WaveFlow provides ops for B-mode image reconstruction (beamforming), signal processing and quantitative ultrasound. The ops were implemented both for the CPU and GPU, as~well~as for built-in automated tests and benchmarks. To demonstrate WaveFlow's performance, ultrasound data were acquired from wire and cyst phantoms and elaborated using selected sequences of the~ops. We implemented and evaluated: Delay-and-Sum beamformer, synthetic transmit aperture imaging (STAI), plane-wave imaging (PWI), envelope detection algorithm and dynamic range clipping. The benchmarks were executed on the NVidia® Titan~X GPU integrated in the USPlatform research scanner (us4us Ltd., Poland). We achieved B-mode image reconstruction frame rates of 55 fps, 17 fps for the STAI and the PWI algorithms, respectively. The results showed the feasibility of real-time ultrasound image reconstruction using WaveFlow operators in the TensorFlow framework. WaveFlow source code can be found at github.com/waveflow-team/waveflow.
\end{abstract}
\IEEEpeerreviewmaketitle

\section{Introduction}
In recent years, we can observe a growing interest in the use of deep learning methods and GPU processing in ultrasound imaging. Various machine learning tasks have been considered for this modality, eg. classification, segmentation (for CAD systems) and image enhancement \cite{classification17byra, segm17smistad, recovry17perdios}. What is more, the implementation of ultrasound imaging pipeline directly on GPU becomes more popular and realistic to perform, mostly due to increasing power of the available technology~\cite{gpu18lewandowski, gpu18dongwoon}. The Software Defined Ultrasound paradigm naturally opens the possibility to apply machine learning tools at any stage of the imaging pipeline, especially to process raw ultrasound radio-frequency (RF) data. This idea is highly promising and becomes well supported by hardware, however the appropriate software architecture still has to be developed. 

Ultrasound image reconstruction algorithms can be defined and implemented as a sequence of signal processing operations like: filtering, Delay-and-Sum (DAS) beamforming, envelope detection and dynamic range clipping. These operations have a clear functional definition and constitute the data processing pipeline. The target implementation of the pipeline should be easy to rearrange depending on the needs, e.g. the software should provide a convenient way to introduce new signal filtering operations. The idea of ultrasound imaging pipeline has been already considered and evaluated in previous publications~\cite{framework12lewandowski, gpu18dongwoon}. Each of this work considers own, different implementation of the pipeline framework, with operations implemented specifically for one type of the processing device. 

In this work we present WaveFlow~\cite{waveflow} -- a set of ultrasound data acquisition and processing tools for TensorFlow. TensorFlow is a framework, which allows to define a dataflow graph, whose nodes represent operators (units of computation, \emph{ops}), and edges represent ops input/output data (tensors)~\cite{tensorflow}. Each op can have an implementation for multiple processing devices, like CPU or GPU. This software gained the most popularity, thanks to a broad library of available machine learning algorithms, however the framework can be used not only for AI training and inference. In WaveFlow, we provide a collection of high-level signal processing ops and tools, which can constitute imaging pipeline. Those operators can be added to the graph, and process RF data to reconstruct B-mode frames. Furthermore, WaveFlow and TensorFlow ops can be placed in the same graph what provides a convenient way to augment ultrasound image reconstruction with machine learning. 

\section{Methodology}

WaveFlow includes ultrasound Environments and signal processing Operators (ops) library for TensorFlow. 

\subsection{Ultrasound Environment}
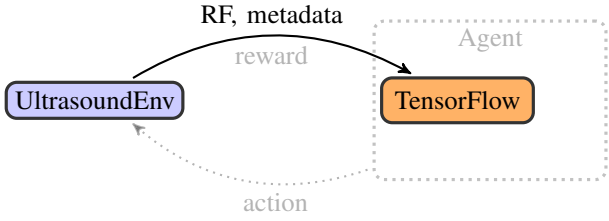
\begin{figure}
\centering
\begin{tikzpicture}[->,node distance=1cm, auto,]
\tikzset{
    agent/.style={
        text width=7.77em,
        minimum height=6em,
    },
}
  \node[node, env](env){UltrasoundEnv};
  \node[node, tf, inner sep=5pt,right=2.6cm of env](graph){TensorFlow};
  \node[node, inactive, minimum height=6em, text width=7.77em, inner sep=5pt, right=2.5cm of env](agent){}; 
  \node[anchor=north, inactive] at (agent.north){Agent}; 
  \path[every node/.style={transform shape, text centered}]
   (env) edge[pil, bend left] node [above] {RF, metadata} node[below, inactive] {reward} (graph)
   (agent) edge[pil, bend left, inactive] node [below, pos=0.37] {action} (env);
\end{tikzpicture}
\caption{Ultrasound Environment provides observations (RF data and metadata), which can be processed by graph defined in TensorFlow. Currently, Ultrasound Environment is an \emph{read-only environment} specialized for ultrasound imaging. }
\label{fig:env}
\end{figure}
The Ultrasound Environment
is a connection point between the input raw RF data source (i.e. dataset, simulation or hardware scanner) and TensorFlow framework. It feeds the graph with:
\begin{itemize}
    \item raw RF data stored as a TensorFlow tensor,
    \item metadata, e.g.:
        \begin{itemize}
            \item examined physical environment parameters, like speed of sound,
            \item ultrasound probe parameters, like aperture size and sampling frequency.
        \end{itemize}
\end{itemize}
The name \emph{Ultrasound Environment} loosely refers to reinforcement learning task environments, where some software agent takes some actions so as to maximize some cumulative reward~\cite{rl10szepesvari}. WaveFlow currently provides only observations from the environment, so it is possible to reconstruct ultrasound image on further steps (see figure~\ref{fig:env}). However, this idea can be extended in future e.g. with actions to perform. 

\subsection{Signal Processing Operators}
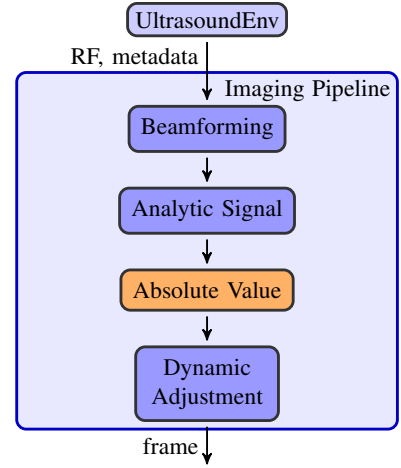
\begin{figure}
\centering
\scalebox{.9}{
\begin{tikzpicture}[->,node distance=1cm, auto,]
  \node[node, env](env){UltrasoundEnv};
  \node[node, draw=blue!77!black, fill=blue!9, very thick, minimum height=15em, text width=15em, inner sep=5pt, below=.5cm of env](pipeline){}; 
  \node[anchor=north east] at (pipeline.north east){Imaging Pipeline}; 
  \node[node, wf, inner sep=5pt,below=1cm of env](bf){Beamforming};
  \node[node, wf, inner sep=5pt,below=.5cm of bf](as){Analytic Signal};
  \node[node, tf, inner sep=5pt,below=.5cm of as](abs){Absolute Value};
  \node[node, wf, inner sep=5pt,below=.5cm of abs, text width=5em](drc){Dynamic Adjustment};
  \node[draw=none] (end) [below=.7cm of drc] {};
  \path[every node/.style={transform shape, text centered}]
   (env) edge[pil] node [left, pos=0.3] {RF, metadata} (bf)
   (bf) edge[pil] node [above] {} (as)
   (as) edge[pil] node [above] {} (abs)
   (abs) edge[pil] node [above] {} (drc)
   (drc) edge[pil] node [left] {frame} (end);
\end{tikzpicture}
}
\caption{Pipeline illustrating B-mode image reconstruction. Blue boxes represent operators defined in WaveFlow, orange -- TensorFlow.}
\label{fig:reconstruction}
\end{figure}

\begin{figure}
\centering
\scalebox{.9}{
\begin{tikzpicture}[->,node distance=1cm, auto,]
  \node[node, env](env){UltrasoundEnv};
  \node[node, wf, inner sep=5pt,below=.5cm of env](bf){Beamforming};
  \node[node, wf, inner sep=5pt,below=.5cm of bf](as){Analytic Signal};
  \node[node, tf, inner sep=5pt,below=.5cm of as](abs){Absolute Value};
  \node[node, draw=blue!77!black, fill=blue!9, very thick, minimum height=11em, text width=14em, inner sep=5pt, right=.2cm of abs](estimator){}; 
  \node[anchor=north east] at (estimator.north east){Estimator}; 
  \node[node, wf, inner sep=5pt,right=.3cm of abs](slice){Slice};
  \node[node, wf, inner sep=4pt,above right=.5cm of slice](moment){$E[X]$};
  \node[node, wf, inner sep=4pt,right=.35cm of slice](moment2){$E[X^2]$};
  \node[node, wf, inner sep=4pt,below right=.5cm of slice](moment3){$E[X^3]$};
  \node[node, tf, inner sep=5pt,right=2.1cm of slice](fc1){FC};
  \node[draw=none, right=.5cm of fc1](ppp){...};
  \node[draw=none] (end2) [right=1.1cm of ppp] {};
  \node[node, wf, inner sep=5pt,below=.5cm of abs, text width=5em, inactive-pipeline](drc){Dynamic Adjustment};
  \node[draw=none] (end) [below=.7cm of drc] {};
  \path[every node/.style={transform shape, text centered}]
   (env) edge[pil] node [left] {RF, metadata} (bf)
  
   (bf) edge[pil] node [above] {} (as)
   (as) edge[pil] node [above] {} (abs)
   (abs) edge[pil] node [above] {} (slice)
   (slice) edge[pil] node [above] {} (moment)
   (slice) edge[pil] node [above] {} (moment2)
   (slice) edge[pil] node [above] {} (moment3)
   (moment) edge[pil] node [above] {} (fc1)
   (moment2) edge[pil] node [above] {} (fc1)
   (moment3) edge[pil] node [above] {} (fc1)
   (fc1) edge[pil] node [above] {} (ppp)
   (ppp) edge[pil] node [above] {[(u,k)]} (end2)
   (abs) edge[pil, inactive-pipeline] node [above, inactive-pipeline] {} (drc)
   (drc) edge[pil, inactive-pipeline] node [left, inactive-pipeline] {frame} (end);
\end{tikzpicture}
}
\caption{Example use case: we can estimate the u and k parameters of the Homodyned~K distribution, which is widely used to model backscattered echo statistics. 
The estimation can be performed using fully connected multi-layer neural network. As the result, quantitative parametric maps of distribution's parameters are obtained. 
Blue boxes represent operators defined in WaveFlow, orange -- TensorFlow.}
\label{fig:estimator}
\end{figure}
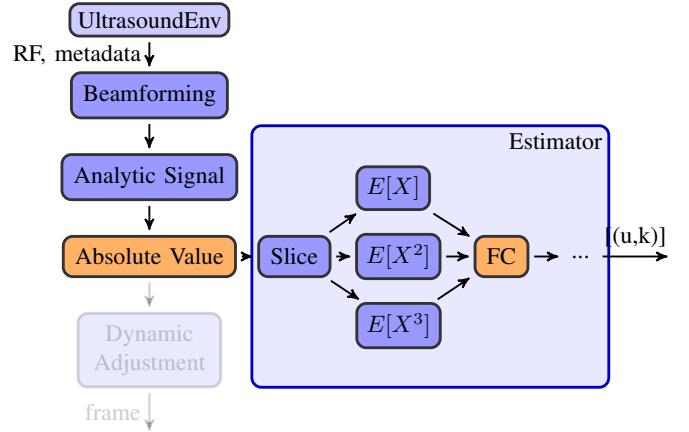

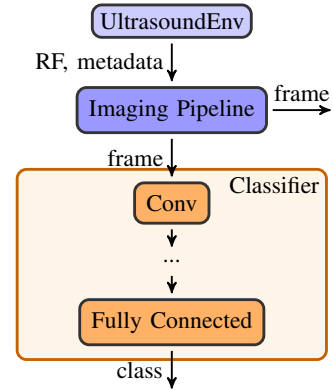
\begin{figure}
\centering
\scalebox{.9}{
\begin{tikzpicture}[->,node distance=1cm, auto,]
  \centering
  \node[node, env](env){UltrasoundEnv};
  \node[node, wf, inner sep=5pt,below=.7cm of env](pipeline){Imaging Pipeline};
  \node[draw=none] (end1) [right=1cm of pipeline] {};
  \node[node, draw=orange!77!black, fill=orange!9, very thick,, minimum height=8em, text width=12em, inner sep=5pt, below=.5cm of pipeline](classifier){}; 
  \node[anchor=north east] at (classifier.north east){Classifier}; 
  \node[node, tf, inner sep=5pt,below=.7cm of pipeline](conv1){Conv};
  \node[draw=none, below=.4cm of conv1](ppp){...};
  \node[node, tf, inner sep=5pt,below=.4cm of ppp](fc2){Fully Connected};
  \node[draw=none] (end2) [below=.7cm of fc2] {};
  \path[every node/.style={transform shape, text centered}]
   (env) edge[pil] node [left] {RF, metadata} (pipeline)
   (pipeline) edge[pil] node [above] {frame} (end1)
   (pipeline) edge[pil] node [left] {frame} (conv1)
   (conv1) edge[pil] node [] {} (ppp)
   (ppp) edge[pil] node [] {} (fc2)
   (fc2) edge[pil] node [left] {class} (end2)
   ;
\end{tikzpicture}
}
\caption{Example use case: we can classify B-mode frame using convolutional neural network. 
As the result, we obtain B-mode frame and its class. Blue boxes represent operators defined in WaveFlow, orange -- TensorFlow.}
\label{fig:classifier}
\end{figure}

The data provided by the Ultrasound Environment can be processed by operators registered in TensorFlow framework. WaveFlow includes implementations of the following ops:
\begin{itemize}
    \item pre-processing: e.g. FIR filter;
    \item mid-processing: Delay-and-Sum (DAS) beamformer, including the Synthetic Transmit Aperture Imaging (STAI) and the Plane Wave Imaging (PWI);
    \item post-processing: e.g. envelope detection, dynamic adjustment.
\end{itemize}
These ops can be chained into a classical ultrasound image reconstruction pipeline, as presented in figure~\ref{fig:reconstruction}. The pipeline can be augmented with machine learning tools at any step of processing. Here we present two simple use cases as an example: applying neural network to estimate homodyned~K distribution parameters from beamformed data (figure~\ref{fig:estimator}), and applying convolutional network to classify beamformed image (figure~\ref{fig:classifier}).
Our public library of ops will be extended with the quantitative ultrasound (qUS) estimators in future.

Whenever possible, convenient and efficient, we implemented WaveFlow operations as a composition of ops available in the TensorFlow core library (e.g. Hilbert transform using the FFT). For other cases (e.g. DAS) we have provided our own implementations for the CPU and GPU.   

\section{Evaluation}

The most recent performance evaluations are available at~\cite{waveflow}. Here we present the imaging pipeline throughput (figure~\ref{fig:reconstruction}), evaluated using the following software and hardware:
\begin{itemize}
    \item WaveFlow ver.~0.1 (March 28, 2018);
    \item TensorFlow ver.~1.6.0;
    \item CUDA ver.~9.0;
    \item NVidia Titan~X GPU integrated in the~USPlatform research scanner (us4us Ltd., Poland).
\end{itemize}

Experiments were performed according to the following assumptions:
\begin{itemize}
    \item STAI: input tensor with a shape (128, 64, 2048) [number of acquisitions, number of channels, number of samples] per frame, data type: float64, output frame: (2048, 128), single focal point, wire/cyst phantom data examined, available here~\cite{usgdata};
    \item PWI: input tensor with a shape (11, 192, 2048), data type: float64, output frame (512, 128), wire/cyst phantom data examined, available here~\cite{usgdata};
    \item dynamic clipped to range [0, 30] dB.
\end{itemize}

\begin{table}[]
\caption{Results achieved on each step of ultrasound imaging pipeline.}
\label{table:results}
\centering
\begin{tabular}{|c|c|c|}
\hline
\rowcolor[HTML]{C0C0C0} 
\textbf{Step}          & \textbf{STAI {[}ms/frame{]}} & \multicolumn{1}{l|}{\cellcolor[HTML]{C0C0C0}\textbf{PWI {[}ms/frame{]}}} \\ \hline
Beamforming            & 11                          & 55                                                                       \\ \hline
Envelope Detection     & 5                           & 4                                                                        \\ \hline
Dynamic Adjustment & 2                           & 1                                                                        \\ \hline
\end{tabular}
\end{table}

Table~\ref{table:results} presents the number of miliseconds required to process an input data frame at each step of the pipeline. In total, 18~ms/frame was required for the STAI (55~FPS), and 60~ms/frame (17~FPS) for the PWI.

\section{Conclusion}

In this work, we implemented and verified the feasibility and efficiency of WaveFlow --
a set of ultrasound data acquisition and processing tools for TensorFlow on a GPU-based ultrasonic research scanner. 
The results of the experiments show, that our software can be successfully used to reconstruct ultrasound B-mode images in real-time.

Tight integration with TensorFlow opens a possibility to conveniently integrate the imaging pipeline with machine learning algorithms.
It also supports execution on multiple processing units, for the price of accepting limitations of this software (e.g. static dataflow graph). 
WaveFlow extends the TensorFlow framework with a collection of general purpose, signal processing algorithms. 
It can be used to reconstruct ultrasound images in particular, and can be executed on GPU or CPU, 
in experimental real-time environment or for research purposes.

Waveflow is an open-source effort (github.com/waveflow-team/waveflow) and contributors are welcome.

\vfill\null

\end{document}